\documentclass[aps,nofootinbib,amsmath,prd,twocolumn,showpacs,superscriptaddress,groupedaddress]{revtex4-1}

\usepackage[american]{babel}
\usepackage{amsfonts}
\usepackage{amsmath}
\usepackage{amssymb}
\usepackage{slashed}
\usepackage{xcolor}
\usepackage{bm}
\usepackage{float}
\usepackage[utf8]{inputenc}
\usepackage[macrosonly]{chet}
\usepackage{nicefrac}
\usepackage{hyperref}
\usepackage{url}
\usepackage{array}
\usepackage{multirow}
\usepackage{comment}

\usepackage{mathtools,tensor,mhchem}

\usepackage{lmodern}
\usepackage{tikz}
\usetikzlibrary{decorations.pathmorphing}
\usetikzlibrary{positioning}

\tikzset{graviton/.style={decorate, decoration={snake, segment length=2mm, amplitude=0.6mm}}}

\begin{document}

\title{Gradient properties of $\varphi^3$ in $d=6-\varepsilon$}

\author{Lorenzo Benfatto}
\email{lorenzo.benfatto@phd.unipi.it}
\affiliation{Universit\`a di Pisa and INFN - Sezione di Pisa, Largo Bruno Pontecorvo 3, 56127 Pisa, Italy}

\author{Omar Zanusso}
\email{omar.zanusso@unipi.it}
\affiliation{Universit\`a di Pisa and INFN - Sezione di Pisa, Largo Bruno Pontecorvo 3, 56127 Pisa, Italy}

\begin{abstract}
The renormalization group flow of the multiscalar interacting $\varphi^3$ theory in $d=6$ dimensions is known to have a gradient structure, in which suitable generalizations of the beta functions $B^{I}$ emerge as the gradient of a scalar function $A$, $\partial_I A = T_{IJ} B^J $, with a nontrivial tensor $T_{IJ}$ in the space of couplings.
This has been shown directly to three loops in schemes such as $\overline{\rm MS}$ and can be argued in general by identifying $A$ with the coefficient of the topological term of the trace-anomaly in $d=6$ up to a normalization.
In this paper we show that the same renormalization group has a gradient structure in $d=6-\varepsilon$.
The requirement of a gradient structure is translated to linear constraints that the coefficients of the $\overline{\rm MS}$ beta functions must obey, one of which is new and pertinent only to the extension to $d \neq 6$.
%
\end{abstract}

\maketitle

\section{Introduction}\label{sect:intro}

The scalar theory with cubic-power interactions and arbitrary number of flavors described by the action
\begin{equation}\label{eq:model}
 \begin{split}
  S[\varphi] = \int {\rm d}^dx \Bigl\{ \frac{1}{2} \partial_\mu \varphi_i \partial^\mu \varphi_i + \frac{1}{3!} g_{ijk} \varphi_i\varphi_j\varphi_k\Bigr\}
 \end{split}
\end{equation}
is perturbatively renormalizable in $d=6-\varepsilon$ dimensions \cite{Fisher:1978pf,deAlcantaraBonfim:1980pe}.
A summation convention is implied over repeated flavor indices $i=1,\cdots, N$, where $N$ is the total number of scalars.
The renormalization group functions of \eqref{eq:model} are known in complete generality up to three loops \cite{Fei:2014yja,Gracey:2015fia}. These functions include the (symmetric) gamma function $\gamma_{ij}$
coming from the wavefunction renormalization and the beta function $\beta^I$ of the tensor coupling $g^I = g_{ijk}$, where we use the collective index $I$ to denote a fully symmetric triple $(ijk)$.\footnote{%
We choose the collective index $I$ to be covariant because the $g$s are later interpreted
as coordinates in the space of couplings. $I$ spans over $\binom{N+2}{3}$ independent indices.
}

We rescale the couplings $g_{ijk} \to (4\pi)^3 g_{ijk}$ in the original action \eqref{eq:model} to avoid unintresting numerical factors.
The general structure of the beta function in the modified minimal subtraction scheme ($\overline{\rm MS}$) up to any given order of the loop expansion in $d=6$ is
\begin{equation}\label{eq:general-structure}
 \begin{split}
 \beta_{ijk} &= {\cal S}_3\left\{
 \begin{tikzpicture}[scale=0.5,rotate=180,baseline=(vert_cent.base)]
	\node (vert_cent) at (0,0) {$\phantom{\cdot}$};
	\node (center) at (0,0) {};
	\def\radius{1cm};
	\node[inner sep=0pt] (left) at (180:\radius) {};
        \node[inner sep=0pt] (extleft) at (180:1.7*\radius) {};
	\node[inner sep=0pt] (right) at (0:\radius) {};
        \node[inner sep=0pt] (extright) at (0:1.7*\radius) {};
        \node[inner sep=0pt] (extrightup) at (1.7*\radius+0.5*\radius,0.86*\radius) {};
        \node[inner sep=0pt] (extrightdown) at (1.7*\radius+0.5*\radius,-0.86*\radius) {};
        \draw (center) circle[radius=\radius];
	\draw (extright) -- (extleft);
        \draw (extright) -- (extrightup);
        \draw (extright) -- (extrightdown);
        \fill[gray] (center) circle[radius=\radius];
        \filldraw (extright) circle[radius=2pt];
  \end{tikzpicture}
  \right\}
  +
  \begin{tikzpicture}[scale=0.5,baseline=(vert_cent.base)]
	\node (vert_cent) at (0,0) {$\phantom{\cdot}$};
	\node (center) at (0,0) {};
	\def\radius{1cm};
	\node[inner sep=0pt] (topleft) at (120:\radius) {};
        \node[inner sep=0pt] (exttopleft) at (120:1.7*\radius) {};
        \node[inner sep=0pt] (bottomleft) at (240:\radius) {};
        \node[inner sep=0pt] (extbottomleft) at (240:1.7*\radius) {};
	\node[inner sep=0pt] (right) at (0:\radius) {};
        \node[inner sep=0pt] (extright) at (0:1.7*\radius) {};
	\draw (center) circle[radius=\radius];
	\draw (center) -- (exttopleft);
	\draw (center) -- (extbottomleft);
	\draw (center) -- (extright);
        \fill[gray] (center) circle[radius=\radius];
  \end{tikzpicture}\,,
 \end{split}
\end{equation}
where the filled-in circles represent one-particle irreducible contractions of $g_{ijk}$ with no tradpoles. The first term contains the symmetric $\gamma$ function and it is symmetrized as
${\cal S}_3\left(\gamma_{mi} \, g_{mjk}\right)=\gamma_{mi} \, g_{mjk}+\gamma_{mj}g_{imk}+\gamma_{mk} \, g_{ijm}$,
using the symmetry factor as label.

The analysis of the conformal anomaly of \eqref{eq:model} can be carried out using local renormalization group methods of Ref.~\cite{Grinstein:2013cka,Poole:2018ljd}. It proves that in $d=6$ there exists a special scalar function of the couplings, $A=A(g)$, roughly related to the topological anomaly \cite{Grinstein:2015ina,Stergiou:2016uqq}, such that its gradient has the form
\begin{equation}\label{eq:gradient}
 \begin{split}
  \partial_I A = T_{IJ} B^J
  \,,
 \end{split}
\end{equation}
which we also refer to as integrable.
On the rhs of \eqref{eq:gradient} we introduced a tensor in the space of couplings, $T_{IJ}$,
and the $B$-functions, that are related to the $\beta$-functions by
\begin{equation}\label{eq:B-functions}
 \begin{split}
  B^I = \beta^I - (\upsilon \cdot g)^I
  \,.
 \end{split}
\end{equation}
The tensor $\upsilon_{ij} = \upsilon_{[ij]}$ should be interpreted as the antisymmetric part of the gamma functions, which acts on $g^I$ as $(\upsilon \cdot g)^I = \upsilon_{mi} \, g_{mjk}+\upsilon_{mj} \, g_{imk}+\upsilon_{mk} \, g_{ijm}$,
i.e., as an element of the orthogonal algebra. The complete matrix $(\gamma+\upsilon)_{ij}$ can be obtained by renormalizing a local source term for $\varphi^i$ rather than the two-point functions \cite{Herren:2021yur}.

The condition $B^I=0$, rather than $\beta^I=0$, is the necessary requirement for conformal invariance, in that the operatorial relation $[T^\mu{}_\mu] = B^I [{\cal O}_I] $ 
holds on-shell and in the flat space limit for the renormalized energy-momentum tensor $T_{\mu\nu}$ and the renormalized operators ${\cal O}_I = \varphi^3_I$ ($T_{\mu\nu}$ should not be confused with the tensor $T_{IJ}$ in the space of couplings) as a consequence of the Ward identities of curved space's Weyl transformations that are improved by a scale dependent flavor rotation \cite{Jack:2013sha}. The ``improvement'' is precisely the second term of Eq.~\eqref{eq:B-functions}. Fixed points have been classified, at least numerically, for low values of $N$, see e.g., Refs.~\cite{Osborn:2017ucf,Osborn:2020cnf,Codello:2019isr,Safari:2020eut}.

The existence of $A$ is particularly useful when trying to establish a perturbative proof of the strong version of the $a$-theorem, i.e., the irreversibility of the renormalization group. For example, under the above assumptions the flow of $A$ generated by the $B$s becomes
\begin{equation}\label{eq:irreversibility}
 \begin{split}
  B^I\partial_I A = G_{IJ} B^I B^J
  \,,
 \end{split}
\end{equation}
in which we defined $G_{IJ} = T_{(IJ)}$. Then $B^I\partial_I A$ is always positive if $G_{IJ}$ is positive and nondegenerate, i.e., if $G_{IJ}$ is an Euclidean metric in the space of couplings. The latter fact can generally be established, at least perturbatively, for unitary theories and has lead to generalizations of Zamolodchikov's theorem \cite{Zamolodchikov:1986gt} to $d=4$ dimensions \cite{Osborn:1991gm}. In $d=2$ it can be shown explicitly that there exists a scheme for which $G_{IJ}$ is precisely Zamolodchikov's metric \cite{Osborn:1991gm}.

In arbitrary $d$ one may instead imagine to rely on the generalization of $A$ to the $F$-function, that is, to the free energy of CFTs evaluated on spheres up to a normalization, defined $\tilde{F} = \sin(\pi d/2) \log Z_{S^d}$ \cite{Fei:2015oha,Fei:2014yja}, which interpolates with the topological anomalies in even dimensions and is also believed to be a monotonic function of the renormalization group. This suggests that Eq.~\eqref{eq:gradient} may also hold in the $\varepsilon$ expansion when $d=6-\varepsilon$ and the interaction becomes slightly relevant.
However, for this logic to be sound, the identification must also hold slightly away from the CFTs critical values.
This possibility has already been confirmed for $\varphi^4$ in $d=4-\varepsilon$ in Ref.~\cite{Pannell:2024sia}, but the $\varphi^3$ case considered here is still unknown.

The gradient structure \eqref{eq:gradient} of the model \eqref{eq:model} has already been verified to three loops in $d=6$ some time ago (in two schemes, including $\overline{\rm MS}$) \cite{Gracey:2015fia}.
The requirement of integrability can be formulated as independent linear constraints that
the coefficients of the beta function of higher loops should satisfy. Specifically there is one constraint for the coefficients at two loops and ten constraints for those at three loops.
In this paper we show that in $d=6-\varepsilon$ there is an additional constraint coming from the three loop order of Eq.~\eqref{eq:gradient}. This constraint is independent from the others and verified by the actual $\overline{\rm MS}$ coefficients, which gives further evidence of the link between the $A$ and $F$ functions and opens the avenue to a perturbative proof of the $F$-theorem for the cubic model in the spirit of Ref.~\cite{Pannell:2025ixz}.

\section{Loop expansion}\label{sect:loop}

We first recall, as briefly as possible, the results in $d=6$ up to three loops in order to set the notation before moving to the $\varepsilon$-expansion in the next section. We represent contractions of the tensor couplings $g_{ijk}$ diagrammatically:
to each black dot corresponds an insertion of the coupling tensor and to each connecting line a contracted index.
At one loop the symmetric part of the anomalous dimension is
\begin{equation}\label{eq:gamma1}
 \begin{split}
  \gamma^{(1)}_{ij} &= c_{1,a}\,
  \begin{tikzpicture}[scale=0.5,baseline=(vert_cent.base)]
	\node (vert_cent) at (0,0) {$\phantom{\cdot}$};
	\node (center) at (0,0) {};
	\def\radius{1cm};
	\node[inner sep=0pt] (left) at (180:\radius) {};
        \node[inner sep=0pt] (extleft) at (180:1.7*\radius) {};
	\node[inner sep=0pt] (right) at (0:\radius) {};
        \node[inner sep=0pt] (extright) at (0:1.7*\radius) {};
	\draw (center) circle[radius=\radius];
	\filldraw (left) circle[radius=2pt];
	\filldraw (right) circle[radius=2pt];
	\draw (left) -- (extleft);
	\draw (right) -- (extright);
  \end{tikzpicture}
 \end{split}
\end{equation}
and there is no antisymmetric tensor $\upsilon_{ij}$ because it is not possible to form an antisymmetric combination with two couplings.
The beta function is
\begin{equation}\label{eq:beta1}
 \begin{split}
  \beta^{(1)}_{ijk} &=
  {\cal S}_3\left(\gamma^{(1)}_{im}g_{mjk}\right)
  +b_{1,a}\,
  \begin{tikzpicture}[scale=0.5,baseline=(vert_cent.base)]
	\node (vert_cent) at (0,0) {$\phantom{\cdot}$};
	\node (center) at (0,0) {};
	\def\radius{1cm};
	\node[inner sep=0pt] (topleft) at (120:\radius) {};
        \node[inner sep=0pt] (exttopleft) at (120:1.7*\radius) {};
        \node[inner sep=0pt] (bottomleft) at (240:\radius) {};
        \node[inner sep=0pt] (extbottomleft) at (240:1.7*\radius) {};
	\node[inner sep=0pt] (right) at (0:\radius) {};
        \node[inner sep=0pt] (extright) at (0:1.7*\radius) {};
	\draw (center) circle[radius=\radius];
	\filldraw (topleft) circle[radius=2pt];
        \filldraw (bottomleft) circle[radius=2pt];
	\filldraw (right) circle[radius=2pt];
	\draw (topleft) -- (exttopleft);
	\draw (bottomleft) -- (extbottomleft);
	\draw (right) -- (extright);
  \end{tikzpicture}\,,
 \end{split}
\end{equation}
which generally depends on the two parameters $c_{1,a}$ and $b_{1,a}$. The $\overline{\rm MS}$ values are $c_{1,a}=\frac{1}{12}$ and $b_{1,a}=-1$. Given $\upsilon_{ij}^{(1)}=0$, there is no difference between beta- and $B$-functions at one loop, $\beta^{(1)} =B^{(1)}$.
The $A$ function is parametrized by irreducible bubbles as
\begin{equation}\label{eq:afun1}
 \begin{split}
 A^{(1)} &= 
 a_{1,a}\,
  \begin{tikzpicture}[scale=0.5,baseline=(vert_cent.base)]
	\node (vert_cent) at (0,0) {$\phantom{\cdot}$};
	\node (center) at (0,0) {};
	\def\radius{1cm};
	\node[inner sep=0pt] (tr) at (45:\radius) {};
	\node[inner sep=0pt] (tl) at (135:\radius) {};
	\node[inner sep=0pt] (bl) at (225:\radius) {};
        \node[inner sep=0pt] (br) at (315:\radius) {};
	\draw (center) circle[radius=\radius];
	\filldraw (tr) circle[radius=2pt];
        \filldraw (tl) circle[radius=2pt];
        \filldraw (bl) circle[radius=2pt];
        \filldraw (br) circle[radius=2pt];
	\draw (tr) -- (bl);
	\draw (tl) -- (br);
	\end{tikzpicture}
+a_{1,b}\,
  \begin{tikzpicture}[scale=0.5,baseline=(vert_cent.base)]
	\node (vert_cent) at (0,0) {$\phantom{\cdot}$};
	\node (center) at (0,0) {};
	\def\radius{1cm};
	\node[inner sep=0pt] (tr) at (45:\radius) {};
	\node[inner sep=0pt] (tl) at (135:\radius) {};
	\node[inner sep=0pt] (bl) at (225:\radius) {};
        \node[inner sep=0pt] (br) at (315:\radius) {};
	\draw (center) circle[radius=\radius];
	\filldraw (tr) circle[radius=2pt];
        \filldraw (tl) circle[radius=2pt];
        \filldraw (bl) circle[radius=2pt];
        \filldraw (br) circle[radius=2pt];
	\draw (tr) to[out=225, in=135] (br);
	\draw (tl) to[out=-45, in=45] (bl);
    \end{tikzpicture}
 \,.
 \end{split}
\end{equation}
The leading order of $T_{IJ}$ is necessarily symmetric, so $T^{(0)}_{IJ} = G^{(0)}_{IJ}$, and is chosen to be trivial, $G^{(0)}_{IJ} = \delta_{IJ}$,
thanks also to the fact that Eq.~\eqref{eq:gradient} is invariant under a combined rescaling of $A$ and $G_{IJ}$ by a constant.
The expansion of Eq.~\eqref{eq:gradient} to order $g^3$ gives $2$ equations,
\begin{equation}\label{eq:one-loop-gradient-constraints}
 \begin{split}
  4 a_{1,a}-b_{1,a}=0\,, \qquad 4 a_{1,b}-3 c_{1,a}=0 \,,
 \end{split}
\end{equation}
which can always be solved for the coefficients of $A^{(1)}$, so there is no constraint from integrability. The quite simple solution is given in Eq.~\eqref{eq:Afun-values1} of Appendix~\ref{sect:constraints}.

At two loops there are $2$ contractions contributing to the anomalous dimension
\begin{equation}\label{eq:gamma2}
 \begin{split}
  \gamma^{(2)}_{ij} &=
  c_{2,a}\,
  \begin{tikzpicture}[scale=0.5,baseline=(vert_cent.base)]
	\node (vert_cent) at (0,0) {$\phantom{\cdot}$};
	\node (center) at (0,0) {};
	\def\radius{1cm};
	\node[inner sep=0pt] (l) at (180:\radius) {};
        \node[inner sep=0pt] (el) at (180:1.7*\radius) {};
	\node[inner sep=0pt] (r) at (0:\radius) {};
        \node[inner sep=0pt] (er) at (0:1.7\radius) {};
	\node[inner sep=0pt] (t) at (90:\radius) {};
        \node[inner sep=0pt] (b) at (-90:\radius) {};
	\draw (center) circle[radius=\radius];
	\filldraw (l) circle[radius=2pt];
        \filldraw (r) circle[radius=2pt];
        \filldraw (t) circle[radius=2pt];
        \filldraw (b) circle[radius=2pt];
        \draw (t) -- (b);
	\draw (l) -- (el);
	\draw (r) -- (er);
 \end{tikzpicture}
  +c_{2,b}\,
  \begin{tikzpicture}[scale=0.5,baseline=(vert_cent.base)]
	\node (vert_cent) at (0,0) {$\phantom{\cdot}$};
	\node (center) at (0,0) {};
	\def\radius{1cm};
	\node[inner sep=0pt] (l) at (180:\radius) {};
        \node[inner sep=0pt] (el) at (180:1.7*\radius) {};
	\node[inner sep=0pt] (r) at (0:\radius) {};
        \node[inner sep=0pt] (er) at (0:1.7\radius) {};
	\node[inner sep=0pt] (tl) at (120:\radius) {};
        \node[inner sep=0pt] (tr) at (60:\radius) {};
	\draw (center) circle[radius=\radius];
	\filldraw (l) circle[radius=2pt];
        \filldraw (r) circle[radius=2pt];
        \filldraw (tl) circle[radius=2pt];
        \filldraw (tr) circle[radius=2pt];
        \draw (r) -- (er);
	\draw (l) -- (el);
	\draw (tl) to[out=120-180, in=60-180] (tr);
\end{tikzpicture}
 \end{split}
\end{equation}
and $3$ additional ones contributing to the beta-function
\begin{equation}\label{eq:beta2}
 \begin{split}
  \beta^{(2)}_{ijk} &=
  {\cal S}_3\left(\gamma^{(2)}_{mi}g_{mjk}\right)
  +b_{2,a}\,{\cal S}_3\,
  \begin{tikzpicture}[scale=0.5,baseline=(vert_cent.base)]
	\node (vert_cent) at (0,0) {$\phantom{\cdot}$};
	\node (center) at (0,0) {};
	\def\radius{1cm};
	\node[inner sep=0pt] (tl) at (120:\radius) {};
        \node[inner sep=0pt] (etl) at (120:1.7*\radius) {};
        \node[inner sep=0pt] (bl) at (240:\radius) {};
        \node[inner sep=0pt] (ebl) at (240:1.7*\radius) {};
	\node[inner sep=0pt] (r) at (0:\radius) {};
        \node[inner sep=0pt] (er) at (0:1.7*\radius) {};
        \node[inner sep=0pt] (itr) at (45:\radius) {};
        \node[inner sep=0pt] (ibr) at (-45:\radius) {};
	\draw (center) circle[radius=\radius];
	\filldraw (tl) circle[radius=2pt];
        \filldraw (bl) circle[radius=2pt];
	\filldraw (r) circle[radius=2pt];
        \filldraw (itr) circle[radius=2pt];
        \filldraw (ibr) circle[radius=2pt];
	\draw (tl) -- (etl);
	\draw (bl) -- (ebl);
	\draw (r) -- (er);
        \draw (itr) to[out=180+45, in=90+45] (ibr);
  \end{tikzpicture}
\\
&
+b_{2,b}\,{\cal S}_3\,
  \begin{tikzpicture}[scale=0.5,baseline=(vert_cent.base)]
	\node (vert_cent) at (0,0) {$\phantom{\cdot}$};
	\node (center) at (0,0) {};
	\def\radius{1cm};
	\node[inner sep=0pt] (tl) at (120:\radius) {};
        \node[inner sep=0pt] (etl) at (120:1.7*\radius) {};
        \node[inner sep=0pt] (bl) at (240:\radius) {};
        \node[inner sep=0pt] (ebl) at (240:1.7*\radius) {};
	\node[inner sep=0pt] (r) at (0:\radius) {};
        \node[inner sep=0pt] (er) at (0:1.7*\radius) {};
        \node[inner sep=0pt] (i1) at (40:\radius) {};
        \node[inner sep=0pt] (i2) at (80:\radius) {};
	\draw (center) circle[radius=\radius];
	\filldraw (tl) circle[radius=2pt];
        \filldraw (bl) circle[radius=2pt];
	\filldraw (r) circle[radius=2pt];
        \filldraw (i1) circle[radius=2pt];
        \filldraw (i2) circle[radius=2pt];
	\draw (tl) -- (etl);
	\draw (bl) -- (ebl);
	\draw (r) -- (er);
        \draw (i1) to[out=40-180, in=80-180] (i2);
  \end{tikzpicture}
+b_{2,c}\,{\cal S}_1\,
  \begin{tikzpicture}[scale=0.5,baseline=(vert_cent.base)]
	\node (vert_cent) at (0,0) {$\phantom{\cdot}$};
	\node (center) at (0,0) {};
	\def\radius{1cm};
	\node[inner sep=0pt] (tl) at (120:\radius) {};
        \node[inner sep=0pt] (etl) at (120:1.7*\radius) {};
        \node[inner sep=0pt] (bl) at (240:\radius) {};
        \node[inner sep=0pt] (ebl) at (240:1.7*\radius) {};
	\node[inner sep=0pt] (r) at (0:\radius) {};
        \node[inner sep=0pt] (er) at (0:1.7*\radius) {};
        \node[inner sep=0pt] (i1) at (240+180:\radius) {};
        \node[inner sep=0pt] (i2) at (120+180:\radius) {};
        \draw [domain=0:120] plot ({cos(\x)*\radius}, {sin(\x)*\radius});
        \draw [domain=240:360] plot ({cos(\x)*\radius}, {sin(\x)*\radius});
	\filldraw (tl) circle[radius=2pt];
        \filldraw (bl) circle[radius=2pt];
	\filldraw (r) circle[radius=2pt];
        \filldraw (i1) circle[radius=2pt];
        \filldraw (i2) circle[radius=2pt];
	\draw (tl) -- (etl);
	\draw (bl) -- (ebl);
	\draw (r) -- (er);
        \draw (i1) -- (bl);
        \draw (i2) -- (tl);
  \end{tikzpicture}
 \end{split}
\end{equation}
The $\overline{\rm MS}$ values are $ c_{2,a}=\frac{1}{18}$, $ c_{2,b}=-\frac{11}{432}$, $ b_{2,a}=-\frac{1}{4}$, $ b_{2,b}=\frac{7}{72}$ and $ b_{2,c}=-\frac{1}{2}$.
As in the one loop case $\upsilon^{(2)}_{ij}=0$ because it is impossible to form antisymmetric contractions also at order $g^4$, which means that $\beta^{(2)}=B^{(2)}$.
There are $5$ irreducible bubbles that contribute to the $A$ function at this order
\begin{equation}\label{eq:afun2}
 \begin{split}
  A^{(2)} &=
  a_{2,a}\,
  \begin{tikzpicture}[scale=0.5,baseline=(vert_cent.base)]
	\node (vert_cent) at (0,0) {$\phantom{\cdot}$};
	\node (center) at (0,0) {};
	\def\radius{1cm};
	\node[inner sep=0pt] (tr) at (30:\radius) {};
        \node[inner sep=0pt] (t) at (90:\radius) {};
        \node[inner sep=0pt] (tl) at (180-30:\radius) {};
        \node[inner sep=0pt] (bl) at (180+30:\radius) {};
	\node[inner sep=0pt] (b) at (-90:\radius) {};
        \node[inner sep=0pt] (br) at (-30:\radius) {};
	\draw (center) circle[radius=\radius];
	\filldraw (tr) circle[radius=2pt];
        \filldraw (t) circle[radius=2pt];
	\filldraw (tl) circle[radius=2pt];
        \filldraw (bl) circle[radius=2pt];
        \filldraw (b) circle[radius=2pt];
        \filldraw (br) circle[radius=2pt];
	\draw (t) -- (b);
	\draw (bl) -- (tr);
	\draw (br) -- (tl);
\end{tikzpicture}
+a_{2,b}\,
    \begin{tikzpicture}[scale=0.5,baseline=(vert_cent.base)]
	\node (vert_cent) at (0,0) {$\phantom{\cdot}$};
	\node (center) at (0,0) {};
	\def\radius{1cm};
	\node[inner sep=0pt] (r) at (0:\radius) {};
        \node[inner sep=0pt] (t) at (120:\radius) {};
        \node[inner sep=0pt] (b) at (-120:\radius) {};
        \node[inner sep=0pt] (ir) at (0:0.5*\radius) {};
	\node[inner sep=0pt] (it) at (120:0.5*\radius) {};
        \node[inner sep=0pt] (ib) at (-120:0.5\radius) {};
	\draw (center) circle[radius=\radius];
        \draw (center) circle[radius=0.5\radius];
	\filldraw (r) circle[radius=2pt];
        \filldraw (t) circle[radius=2pt];
	\filldraw (b) circle[radius=2pt];
        \filldraw (ir) circle[radius=2pt];
        \filldraw (it) circle[radius=2pt];
        \filldraw (ib) circle[radius=2pt];
	\draw (t) -- (it);
	\draw (b) -- (ib);
	\draw (r) -- (ir);
  \end{tikzpicture}
  +a_{2,c}\,
    \begin{tikzpicture}[scale=0.5,baseline=(vert_cent.base)]
	\node (vert_cent) at (0,0) {$\phantom{\cdot}$};
	\node (center) at (0,0) {};
	\def\radius{1cm};
        \node[inner sep=0pt] (l) at (180:\radius) {};
        \node[inner sep=0pt] (il) at (180:0.36\radius) {};
	\node[inner sep=0pt] (tr) at (45:\radius) {};
	\node[inner sep=0pt] (tl) at (135:\radius) {};
	\node[inner sep=0pt] (bl) at (225:\radius) {};
        \node[inner sep=0pt] (br) at (315:\radius) {};
	\draw (center) circle[radius=\radius];
	\filldraw (tr) circle[radius=2pt];
        \filldraw (tl) circle[radius=2pt];
        \filldraw (bl) circle[radius=2pt];
        \filldraw (br) circle[radius=2pt];
        \filldraw (l) circle[radius=2pt];
        \filldraw (il) circle[radius=2pt];
	\draw (tr) to[out=225, in=135] (br);
	\draw (tl) to[out=-45, in=90] (il);
        \draw (il) to[out=-90, in=45] (bl);
        \draw (l) -- (il);
    \end{tikzpicture}
\\&
 +a_{2,d}\,
    \begin{tikzpicture}[scale=0.5,baseline=(vert_cent.base)]
	\node (vert_cent) at (0,0) {$\phantom{\cdot}$};
	\node (center) at (0,0) {};
	\def\radius{1cm};
        \node[inner sep=0pt] (t) at (90:\radius) {};
        \node[inner sep=0pt] (b) at (-90:\radius) {};
	\node[inner sep=0pt] (tr) at (45:\radius) {};
	\node[inner sep=0pt] (tl) at (135:\radius) {};
	\node[inner sep=0pt] (bl) at (225:\radius) {};
        \node[inner sep=0pt] (br) at (315:\radius) {};
	\draw (center) circle[radius=\radius];
	\filldraw (tr) circle[radius=2pt];
        \filldraw (tl) circle[radius=2pt];
        \filldraw (bl) circle[radius=2pt];
        \filldraw (br) circle[radius=2pt];
        \filldraw (t) circle[radius=2pt];
        \filldraw (b) circle[radius=2pt];
	\draw (tr) to[out=225, in=135] (br);
	\draw (tl) to[out=-45, in=45] (bl);
        \draw (t) -- (b);
    \end{tikzpicture}
 +a_{2,e}\,
    \begin{tikzpicture}[scale=0.5,baseline=(vert_cent.base)]
	\node (vert_cent) at (0,0) {$\phantom{\cdot}$};
	\node (center) at (0,0) {};
	\def\radius{1cm};
        \node[inner sep=0pt] (v1) at (0:\radius) {};
        \node[inner sep=0pt] (v2) at (60:\radius) {};
	\node[inner sep=0pt] (v3) at (120:\radius) {};
	\node[inner sep=0pt] (v4) at (180:\radius) {};
	\node[inner sep=0pt] (v5) at (240:\radius) {};
        \node[inner sep=0pt] (v6) at (300:\radius) {};
	\draw (center) circle[radius=\radius];
	\filldraw (v1) circle[radius=2pt];
        \filldraw (v2) circle[radius=2pt];
        \filldraw (v3) circle[radius=2pt];
        \filldraw (v4) circle[radius=2pt];
        \filldraw (v5) circle[radius=2pt];
        \filldraw (v6) circle[radius=2pt];
	\draw (v1) to[out=180, in=60-180] (v2);
	\draw (v3) to[out=120-180, in=180-180] (v4);
        \draw (v5) to[out=240-180, in=300-180] (v6);
    \end{tikzpicture}
    \,.
 \end{split}
\end{equation}
A general parametric form of the tensor $T^{(1)}_{IJ}$ is obtained by labelling two independent vertices in the bubbles of the previous order, given in Eq.~\eqref{eq:afun1}, and taking derivatives with respect to them. This results in a symmetric tensor, so $T^{(1)}_{IJ}=G^{(1)}_{IJ}$ parametrized by $4$ terms
\begin{equation}\label{eq:metric1}
 \begin{split}
 G^{(1)}_{IJ} &=
 m_{1,a}\,{\cal S}_{18}\,\left[
    \begin{tikzpicture}[scale=0.5,baseline=(vert_cent.base)]
	\node (vert_cent) at (0,0) {$\phantom{\cdot}$};
	\node (center) at (0,0) {};
	\def\vlength{0.7cm};
        \def\hlength{1cm}
        \node[inner sep=0pt] (l1) at (-1*\hlength,1*\vlength) {};
        \node[inner sep=0pt] (l2) at (-1*\hlength,0*\vlength) {};
	\node[inner sep=0pt] (l3) at (-1*\hlength,-1*\vlength) {};
	\node[inner sep=0pt] (r1) at (1*\hlength,1*\vlength) {};
	\node[inner sep=0pt] (r2) at (1*\hlength,0*\vlength){};
        \node[inner sep=0pt] (r3) at (1*\hlength,-1*\vlength) {};
        \node[inner sep=0pt] (i1) at (0*\hlength,1*\vlength) {};
        \node[inner sep=0pt] (i2) at (0*\hlength,0*\vlength) {};
	\filldraw (i1) circle[radius=2pt];
        \filldraw (i2) circle[radius=2pt];
        \draw (l1) -- (r1);
        \draw (l2) -- (r2);
        \draw (i1) -- (i2);
        \draw (l3) -- (r3);
    \end{tikzpicture}
    \right]
 +m_{1,b}\,{\cal S}_{18}\,\left[
    \begin{tikzpicture}[scale=0.5,baseline=(vert_cent.base)]
	\node (vert_cent) at (0,0) {$\phantom{\cdot}$};
	\node (center) at (0,0) {};
	\def\vlength{0.7cm};
        \def\hlength{1cm}
        \node[inner sep=0pt] (l1) at (-1*\hlength,1*\vlength) {};
        \node[inner sep=0pt] (l2) at (-1*\hlength,0*\vlength) {};
	\node[inner sep=0pt] (l3) at (-1*\hlength,-1*\vlength) {};
	\node[inner sep=0pt] (r1) at (1*\hlength,1*\vlength) {};
	\node[inner sep=0pt] (r2) at (1*\hlength,0*\vlength){};
        \node[inner sep=0pt] (r3) at (1*\hlength,-1*\vlength) {};
        \node[inner sep=0pt] (i1) at (-0.5*\hlength,0*\vlength) {};
        \node[inner sep=0pt] (i2) at (0.5*\hlength,0*\vlength) {};
	\filldraw (i1) circle[radius=2pt];
        \filldraw (i2) circle[radius=2pt];
        \draw (l1) -- (r1);
        \draw (l3) -- (r3);
        \draw (l2) -- (i1);
        \draw (r2) -- (i2);
        \draw (i1) to[out=90, in=90] (i2);
        \draw (i1) to[out=-90, in=-90] (i2);
    \end{tikzpicture}
    \right]
    \\&
    +m_{1,c}\,{\cal S}_9\,\left[
    \begin{tikzpicture}[scale=0.5,baseline=(vert_cent.base)]
	\node (vert_cent) at (0,0) {$\phantom{\cdot}$};
	\node (center) at (0,0) {};
	\def\vlength{0.7cm};
        \def\hlength{1cm}
        \node[inner sep=0pt] (l1) at (-1*\hlength,1*\vlength) {};
        \node[inner sep=0pt] (l2) at (-1*\hlength,0*\vlength) {};
	\node[inner sep=0pt] (l3) at (-1*\hlength,-1*\vlength) {};
	\node[inner sep=0pt] (r1) at (1*\hlength,1*\vlength) {};
	\node[inner sep=0pt] (r2) at (1*\hlength,0*\vlength){};
        \node[inner sep=0pt] (r3) at (1*\hlength,-1*\vlength) {};
        \node[inner sep=0pt] (i1) at (0*\hlength,1*\vlength) {};
        \node[inner sep=0pt] (i2) at (0*\hlength,-1*\vlength) {};
	\filldraw (i1) circle[radius=2pt];
        \filldraw (i2) circle[radius=2pt];
        \draw (l1) -- (r1);
        \draw (l2) to[out=0, in=-90] (i1);
        \draw (i2) to[out=90, in=180] (r2);
        \draw (l3) -- (r3);
    \end{tikzpicture}
    \right]
    +m_{1,d}\,{\cal S}_9\,\left[
    \begin{tikzpicture}[scale=0.5,baseline=(vert_cent.base)]
	\node (vert_cent) at (0,0) {$\phantom{\cdot}$};
	\node (center) at (0,0) {};
	\def\vlength{0.7cm};
        \def\hlength{1cm};
        \node[inner sep=0pt] (l1) at (-1*\hlength,1*\vlength) {};
        \node[inner sep=0pt] (l2) at (-1*\hlength,0*\vlength) {};
	\node[inner sep=0pt] (l3) at (-1*\hlength,-1*\vlength) {};
	\node[inner sep=0pt] (r1) at (1*\hlength,1*\vlength) {};
	\node[inner sep=0pt] (r2) at (1*\hlength,0*\vlength){};
        \node[inner sep=0pt] (r3) at (1*\hlength,-1*\vlength) {};
        \node[inner sep=0pt] (i1) at (-0.5*\hlength,0.5*\vlength) {};
        \node[inner sep=0pt] (i2) at (0.5*\hlength,0.5*\vlength) {};
	\filldraw (i1) circle[radius=2pt];
        \filldraw (i2) circle[radius=2pt];
        \draw (l1) to[out=0, in=90] (i1);
        \draw (l2) to[out=0, in=-90] (i1);
        \draw (i2) to[out=90, in=180] (r1);
        \draw (i2) to[out=-90, in=180] (r2);
        \draw (i1) -- (i2);
        \draw (l3) -- (r3);
    \end{tikzpicture}
    \right]\,.
 \end{split}
\end{equation}
Solving Eq.~\eqref{eq:gradient} to order $g^7$ we find $8$
independent equations, which do not admit a general solution for arbitrary coefficients $b_{2}$s and $c_{2}$s.
By linearly eliminating the $a_2$s and $m_1$s coefficients, there is only one surviving equation involving the coefficients of the beta function,
\begin{equation}\label{eq:gradient-constraint-two-loop-previous-inserted}
 \begin{split}
  b_{2,b}  + 6 c_{2,a} +c_{2,b} =0\,,
 \end{split}
\end{equation}
which should be seen as a linear constraint among the $b_2$s and $c_2$s (here we inserted the values of $b_1$s and $c_1$s of the previous order, the general formula is given in Eq.~\eqref{eq:gradient-constraint-two-loop}). This agrees with Ref.~\cite{Gracey:2015fia}.
We also note that, solving for the $a_2$s and $m_1$s, the solution admits $2$ free parameters, $\alpha_1$ and $\tilde{\alpha}_1$ on which solutions depend explicitly, as can be checked in Eqs.~\eqref{eq:Afun-values2} and \eqref{eq:G-values2} of Appendix~\ref{sect:constraints}. The parameters are also discussed briefly in Sect.~\ref{sect:further}.

\begin{widetext}

The renormalization group function to three loops can be found in Ref.~\cite{Gracey:2015fia}, but also deduced from the tables given in Ref.~\cite{Fei:2014yja} for a special model that can be adapted to the fully general case.
There are in total $9$ contractions contributing to the symmetric anomalous dimension,
\begin{equation}\label{eq:gamma3}
 \begin{split}
  \gamma^{(3)}_{ij} &=
    c_{3,a}\,

    \,.
 \end{split}
\end{equation}
At this order the antisymmetric part of the anomalous is nonzero and we parametrize it as
\begin{equation}\label{eq:upsilon3}
\begin{split}
    \upsilon^{(3)}_{ij} &=
    s_{3,a}\,\left\{
    %
    \right\}\,.
 \end{split}
\end{equation}
The beta function includes $17$ additional diagrams, that are
\begin{equation}\label{eq:beta3}
 \begin{split}
  \beta^{(3)}_{ijk} &=
  {\cal S}_3\left(\gamma^{(3)}_{mi}g_{mjk}\right)
  +b_{3,a}\,{\cal S}_3\,
  %
  \,.
\end{split}
\end{equation}
At this order $\beta^{(3)}\neq B^{(3)}$, and the expression of the latter can be obtained from Eq.~\eqref{eq:beta3} replacing $\gamma \to \gamma+\upsilon$ in the first term.
The $\overline{\rm MS}$ values of the coefficients of $\beta^{(3)}$, $\gamma^{(3)}$ and $\upsilon^{(3)}$ are summarized in Eq.~\eqref{eq:MSbar-3loop-parameters}.
The irreducible bubbles contributing to the parametrization of $A$ are
\begin{equation}\label{eq:afun3}
 \begin{split}
  A^{(3)} &=
  a_{3,a}\,
    %
\,.
\end{split}
\end{equation}
We do not give the expression for $T^{(2)}_{IJ}$ here for the sake of brevity, note however that it is symmetric, $T^{(2)}_{IJ}=G^{(2)}_{IJ}$ and it contains $29$ independent terms that can be obtained analogously to the previous order, i.e., by labelling and cutting two vertices of the diagrams of Eq.~\eqref{eq:afun2}  ($30$ contractions are reported in Ref.~\cite[Table~4]{Gracey:2015fia}, but $2$ are identical).  
Requiring the gradient structure of Eq.~\eqref{eq:gradient} to the order $g^9$ of the expansion, we find $45$ equations among the coefficients of
$B^{(3)}$, $A^{(3)}$ and $G^{(2)}$.
Substituting the results from the previous loops, and linearly eliminating the coefficients of $A^{(3)}$ and $G^{(2)}$, we confirm a system of $10$ independent equations that forms a set of constraints on the coefficients of $B^{(3)}$. The result agrees with Refs.~\cite{Gracey:2015fia, jack-osborn-unpublished-draft} and we give it in Eqs.~\eqref{eq:gradient-constraint-three-loop-previous-inserted} and \eqref{eq:gradient-constraint-three-loop} of Appendix~\ref{sect:constraints}. 
\end{widetext}

\section{Solutions in $d=6-\varepsilon$}\label{sect:epsilon}

In the $\varepsilon$-expansion, i.e., for $d=6-\varepsilon$, the interaction is slightly relevant, now with dimension $\frac{\varepsilon}{2}$.
The beta function of the dimensionless coupling must be complemented by the leading term $\beta^{(0)}_{ijk} = B^{(0)}_{ijk}= -\frac{1}{2}\varepsilon g_{ijk}$, which, in turn, forces the leading term of $A$ to be
\begin{equation}\label{eq:afun0}
 \begin{split}
 A^{(0)} = -\frac{\varepsilon}{4} \,
 \begin{tikzpicture}[scale=0.5,baseline=(vert_cent.base)]
	\node (vert_cent) at (0,0) {$\phantom{\cdot}$};
	\node (center) at (0,0) {};
	\def\radius{1cm};
	\node[inner sep=0pt] (left) at (180:\radius) {};
	\node[inner sep=0pt] (right) at (0:\radius) {};
	\draw (center) circle[radius=\radius];
	\filldraw (left) circle[radius=2pt];
	\filldraw (right) circle[radius=2pt];
	\draw (left) -- (right);
  \end{tikzpicture}\,,
 \end{split}
\end{equation}
which solves the ``zeroth'' order (zeroth order if compared to the loop-expansion).

In general, the leading term $B^{(0)}$ of the beta function conspires in Eq.~\eqref{eq:gradient} with $G^{(n)}$ at the $n$th order to produce terms of the same order as, for example, $G^{(0)}*B^{(n)} = B^{(n)}$.
Expanding Eq.~\eqref{eq:gradient}, we have the schematic structure
\begin{equation}\label{eq:gradient-expanded}
    \begin{split}
    \partial A^{(n)} = \sum_{0\leq p \leq n} T^{(p)} * B^{(n-p)}\,,
    \end{split}
\end{equation}
that must be solved order-by-order also including the term $p=n$, which has not appeared in the previous section. In the applications of this section we always have $n\leq 2$, which ensures $T^{(p)}_{IJ} =G^{(p)}_{IJ}$ for the reasons explained in Sect.~\ref{sect:loop}.
A solution is thus found for $\{A^{(n)}, G^{(n)}\}$, rather than for $\{A^{(n)}, G^{(n-1)}\}$, at the $n$th order of the above expansion given $B^{(n)}$, see Ref.~\cite{Pannell:2024sia}. By construction, in the limit $\varepsilon\to 0$ the loop-perturbative solution \emph{must} be recovered, consequently the coefficients of 
$\{A^{(n)}, G^{(n)}\}$ are determined as a regular series in $\varepsilon$.

Given that, to each order $n$ of the expansion in Eq.~\eqref{eq:gradient-expanded}, the coefficients of the solutions of $A^{(n)}$ and $G^{(n)}$ must themselves be regular series in $\varepsilon$, we parametrize
them as done in the pertinent formulas of the loop expansion of Sect.~\ref{sect:loop}, but replacing
\begin{equation}\label{eq:coefficients-expanded}
    \begin{split}
    a_{n,x}\to \tilde{a}_{n,x}(\varepsilon)\,, \qquad
    m_{n,x} \to \tilde{m}_{n,x}(\varepsilon)\,,
    \end{split}
\end{equation}
and the solutions of the previous sections become boundary conditions,
$\tilde{a}_{n,x}(0)=a_{n,x}$ and $\tilde{m}_{n,x}(0)=m_{n,x}$.
With the above replacement, at any $n$, Eq.~\eqref{eq:gradient-expanded} gives linear relations between $\tilde{a}_{n,x}$ and $\tilde{m}_{n,x}$, in which the latter are multiplied by $\varepsilon$ because of the contraction $G^{(n)}*B^{(0)} \propto \varepsilon\, G^{(n)}$, i.e., the term $p=n$ in the expansion of Eq.~\eqref{eq:gradient-expanded}.

New constraints among the beta function coefficients can emerge by inspecting the order $\varepsilon$ equation of the resulting systems. One way to find such constraints is to solve
for the coefficients $a_{n,x}$ as functions of $(b_n,c_n,s_n)$ and for the coefficients $m_{n,x}$
as functions of $(b_{n+1},c_{n+1},s_{n+1})$, either way the solutions are subject to the constraints of the previous section given in Eqs.~\eqref{eq:gradient-constraint-two-loop-previous-inserted} and \eqref{eq:gradient-constraint-three-loop-previous-inserted}.
Then, the order $n$ of Eq.~\eqref{eq:gradient-expanded} expanded to order $\varepsilon$
relates linear combinations of $\tilde{a}'_{n,x}(0)$ with combinations of $(b_{n+1},c_{n+1},s_{n+1})$ and constraints on the latter set can be found by linearly eliminating
the unknowns $(\tilde{a}'_{n,x}(0),\tilde{m}'_{n,x}(0))$ and inserting the former set for simplicity.
Equivalently, we could solve for the $\tilde{m}_n$ coefficients and demand their regularity, i.e., that there are no $\frac{1}{\varepsilon}$ poles in their series (recall that $G^{(n)}*\beta^{(0)}$ multiplies the  $\tilde{m}_n$s by $\varepsilon$, so their general solution
may contain a pole), as we believe was done in Ref.~\cite{Pannell:2024sia}. Either way, we obtain the same constraints independently of the approach.

At the leading nontrivial order, corresponding to $n=1$ in Eq.~\eqref{eq:gradient-expanded}, there are no additional (independent) constraints, though we observe that the constraint given in Eq.~\eqref{eq:gradient-constraint-two-loop-previous-inserted} is explicitly required for a consistent solution. The values of the leading-order $\varepsilon$ corrections
to the coefficients of $A^{(1)}$ are given in Eq.~\eqref{eq:Afun-values1-epsilon} of Appendix~\ref{sect:constraints}.

At the next-to-leading order, corresponding to $n=2$ in Eq.~\eqref{eq:gradient-expanded},
we find several constraints, most of which linearly related to those given in Appendix~\ref{sect:constraints}.
However, there is one (and only one) new constraint on the coefficients of $B^{(3)}$,
which is independent of the $10$ constraints coming from the loop expansion given in Eq.~\eqref{eq:gradient-constraint-three-loop-previous-inserted}. We find
\begin{equation}\label{eq:gradient-constraint-NLO-new-previous-inserted}
\begin{split}
 &
 36 c_{3,a}-1728 s_{3,a}-144 c_{3,b}+36 b_{3,f}-144 b_{3,g}
 +72 b_{3,h}
 \\&
 -432 c_{3,d}+216 c_{3,e}+5184  c_{3,g}+432 c_{3,c}-7=0
 \,,
\end{split}
\end{equation}
which we see as the main result of this paper. It is a straightforward check that
the above constraint is solved by the $\overline{\rm MS}$ values given in Eq.~\eqref{eq:MSbar-3loop-parameters}. In other words, the integrability of the renormalization group parametrized by the $B^I$ functions in $d=6-\varepsilon$ is not guaranteed by the integrability in $d=6$, instead the coefficients of the three loop beta function must satisfy an additional relation.
The most general form of Eq.~\eqref{eq:gradient-constraint-NLO-new-previous-inserted} in which the previous parameters have not been substituted is presented in Eq.~\eqref{eq:gradient-constraint-NLO-new} of Appendix~\ref{sect:constraints}. The number of coefficients of the various functions and the constraints are summarized in Table~\ref{tab:MScase}.
\begin{table}[htbp]
    \centering
    \begin{tabular}{|c|c|c|c|}
     \hline
     \multicolumn{4}{|c|}{$\overline{\rm MS}^{\phantom{1}}$ renormalization group} \\
     \hline
       Number of terms  & 1 Loop & 2 Loops & 3 Loops  \\
     \hline
        $\beta$ (irr.) & 1 & 3 & 17 \\
        $\gamma$ & 1& 2 & 9 \\
        $\upsilon$ & 0 & 0 & 1 \\
        $B$ (total) & 2 & 5 & 27 \\
        \hline \hline
        $A$ & 2 & 5 & 16 \\
        $G$ &1 & 4 & 29 \\
        \hline \hline
        Loop constraints & 0 & 1 & 10 \\
        $\varepsilon$-expansion constraints & 0 & 0 & 1 \\
        \hline
    \end{tabular}
    \caption{Number of terms and constraints among the tensor structures in the $\overline{\rm MS}$ scheme.}
    \label{tab:MScase}
\end{table}

\section{Further results}\label{sect:further}

The presumed relation between the $d$-dependent $F$-function and the $\varepsilon$-dependent $A$-function has a key role in motivating our original expectation that the flow of the $B$-functions is integrable also in the $\epsilon$-expansion.
In this case it becomes important to assess the role of the tensor $G_{IJ}$ and its interpretation from the point of view of the $F$-theorem. In a recent work, Ref.~\cite{Pannell:2025ixz},
it was shown that, for the $\varphi^4$ theory, it is possible to normalize the quantities and fix parametric dependencies present in $G_{IJ}$ in such a way that it directly generalizes Zamolodchikov's metric in $d\neq 2$, i.e., $G_{IJ} \to C_{IJ} \propto \langle [\varphi^3_I] [\varphi^3_J]\rangle$ where $\varphi^3_I = \varphi_i\varphi_j\varphi_k$ are the interactions and $[\varphi^3_I]$ are the renormalized operators as in Zamolodchikov's paper~\cite{Zamolodchikov:1986gt} and the proportionality factors out the correct power of the geodesic distance between the two operators. The computation of Ref.~\cite{Pannell:2025ixz} was done entirely on spheres, so that the matching extends to the $F$-function. The result is particularly important because in $d=4$ the local renormalization group would instead suggest that $T_{IJ}$ is related to a three-point function, e.g., something like $T_{IJ}\propto \langle [T^\mu{}_\mu][\varphi^3_I] [\varphi^3_J]\rangle$, for which it is difficult to establish a notion of positivity.

In a forthcoming paper we plan to address in more generality many more aspects of the results presented in this paper, including scheme-dependence, but also the geometry of couplings' space.
In anticipation, to spark a discussion, we give the (leading corrections to the) Riemannian curvature scalar ${\cal R}$ associated to $G_{IJ}$, interpreted as a metric. Using
\begin{equation}\label{eq:curvature-expansion}
\begin{split}
 {\cal R} = \partial_I\partial_J G^{(1)}_{IJ}- \partial_I\partial_I G^{(1)}_{JJ}+{\cal O}(g^2) \,,
\end{split}
\end{equation}
we find
\begin{equation}\label{eq:curvature}
\begin{split}
    {\cal R}=&\frac{1}{864} (N+2) N (N-1) \Bigl[
    11N^2+32 N+19
    \\
    &
    +144 (N+3) \tilde{\alpha}_1 
    \Bigr]
    +{\cal O}(\varepsilon,g^2)
    \,,
\end{split}
\end{equation}
which agrees with the result of Ref.~\cite{jack-osborn-unpublished-draft}.
The curvature depends on both the free parameters $\alpha_1 = a_{2,c}$ and
$\tilde{\alpha}_1= m_{1,c} + a_{2,c}$ 
that are not fixed by integrability, see Eqs.~\eqref{eq:Afun-values2} and \eqref{eq:G-values2}
which have been used in Eq.~\eqref{eq:curvature}. These should instead be fixed consistently on spheres by matching $A$ with $\tilde{F}=\sin(\pi d/2) \log Z_{S^d}$ and $G_{IJ}$ with Zamolodchikov's metric under our motivating premise. In the large-$N$ limit we have ${\cal R} \propto N^5$ with a coefficient that
is instead independent on the parameters \cite{jack-osborn-unpublished-draft}, similarly to the result of Ref.~\cite{Pannell:2025ixz}.\footnote{%
We are very grateful to Hugh Osborn for suggestions, for pointing out the mistakes of Eq.~\eqref{eq:curvature} in the previous version of the draft and for kindly sharing updates on Ref.~\cite{jack-osborn-unpublished-draft}. 
}

At subleading orders in the large-$N$ expansion Eq.~\eqref{eq:curvature}
depends on the parameters that are not fixed by the integrability conditions.
As for the interpretation of these parameters,
we observe that $\alpha_1$ is related to the well-known fact that, given a function $A$ that satisfies the integrability condition with a certain metric $G_{IJ}$, we can always find an equivalent function $A' \sim A$ \cite{Osborn:1991gm,Gracey:2015fia},
defined as
\begin{equation}\label{eq:nonuniqueness}
 \begin{split}
  A' = A + d_{IJ} B^I B^J
  \,,
 \end{split}
\end{equation}
for some symmetric matrix $d_{IJ}$, although with a different tensor $T'_{IJ} \sim T_{IJ}$ 
which explicitly is
\begin{equation}\label{eq:nonuniquenessT}
 \begin{split}
  T'_{IJ} = T_{IJ} + {\cal L}_B T_{IJ} + \partial_I(d_{JK} B^K)-\partial_J(d_{IK} B^K)
  \,,
 \end{split}
\end{equation}
where ${\cal L}_B T_{IJ}  = B^K \partial_K T_{IJ} + \partial_I B^K T_{KJ}+ \partial_J B^K T_{IK}$ is the standard Lie derivative.
Notice two important things. On the one hand, at fixed points $g^*_{ijk}$ such that $B^I=0$, we have that $A(g^*)=A'(g^*)$, so the value of $A$ is renormalization group invariant.
On the other hand, if the original $T_{IJ}$ is symmetric (i.e., $T_{IJ}=G_{IJ}$), the equivalence may give an antisymmetric part to $T'_{IJ}$.
By construction, being $G_{IJ}$ the symmetric part of $T_{IJ}$ we always have
\begin{equation}\label{eq:nonuniquenessG}
 \begin{split}
  G'_{IJ} = G_{IJ} + {\cal L}_B d_{IJ} 
  \,,
 \end{split}
\end{equation}
but a symmetric $T_{IJ}$ is mapped to another symmetric $T'_{IJ}$ only if
\begin{equation}
 \begin{split}
  \partial_I(d_{JK} B^K)-\partial_J(d_{IK} B^K)=0\,.
 \end{split}
\end{equation}
This basically is the integrability condition for the $B$s, that is trivially solved by choosing $d_{IJ} = \alpha \, T_{IJ}$ (symmetric) for some constant $\alpha$, in which case the above equation becomes $\left[\partial_I,\partial_J\right]A=0$. Under these assumptions,
\begin{equation}
 \begin{split}
 A' = A + \alpha \, T_{IJ} B^I B^J = A +\alpha\,  B^I \partial_I A\,,
 \end{split}
\end{equation}
so $A'$ would be a renormalization group transformation of $A$ (and obviously still an invariant at fixed points). Notice that $\alpha$ is not necessarily infinitesimal.

In the previous sections we have seen that $T_{IJ}$ is symmetric up to three loops.
In our notation the parameter $\alpha_1$, that does not appear in the scalar curvature, emerges at two loops. In fact
\begin{equation}
 \begin{split}
  A^{(2)}|_{{\rm linear \, in \, }\alpha_1 } 
  &= -2 \alpha_1 G^{(0)}_{IJ} B^{(1)I} B^{(1)J}
  \,,
 \end{split}
\end{equation}
which tells us how to understand the parametric dependence of our result on $\alpha_1$
in relation with $\alpha$. For example, if $A'\equiv A|_{\alpha_1=0}$, then the general $A$ that we have found and given in Eqs.~\eqref{eq:Afun-values1} is such that $A\sim A'=A+\alpha\, G_{IJ}B^I B^J$ for $\alpha=2\alpha_1$.
The dependence on $\tilde{\alpha}_1$ is instead entirely confined to the metric
and can be understood as a consequence of the fact that integrability only fixes the linear combination of the parameters $m_{1,c}+m_{1,d}$, in agreement with the result of Ref.~\cite{Gracey:2015fia}.

Taking all this into account, we can interpret the scalar curvature given in Eq.~\eqref{eq:curvature} as a consequence of the gradient structure of theory space
that demands a nontrivial metric in the space of couplings. However, since the curvature
depends on $\tilde{\alpha}_1$ in a way that is not proportional to the $B$-functions at subleading orders in the large-$N$ expansion, it is not straightforward
to argue for its physical meaning.
However, it may acquire a physical meaning once the parametric ambiguities in both the $A$ function and the metric are resolved through a physically motivated matching procedure,
which would result in matching the values of $\alpha_1$ and $\tilde{\alpha}_1$
by computing $\tilde{F}$ and Zamolodchikov's metric with covariant diagrams on spheres, as discussed earlier.

\section{Conclusions}\label{sect:conclusions}

We have proven that the $\overline{\rm MS}$ renormalization group flow of a multiscalar $\varphi^3$ model
is integrable in $d=6-\varepsilon$ dimensions, i.e., it admits a generating scalar function $A$ at and, importantly, slightly below the upper critical dimension of the field theory,
the latter being the main new result of this paper.
This result is nontrivial because the existence of a scalar $A$ function, dependent on the couplings $g_{ijk}$, is generally guaranteed
by the analysis of the topological term of the conformal anomaly only in $d=6$, i.e., only at the upper critical dimension of the $\varphi^3$ model and not below it.

In general $d$ one could argue that the existence of the so-called CFT $F$-function (the free energy of the CFT on a sphere up to a $d$-dependent constant), which is believed to replace $A$ and interpolate with the topological anomaly itself
in even dimensions by analytic continuation,
should motivate the existence of some $A$ in general $d$ and, specifically, in $d=6-\epsilon$.
However this may not be obvious because the $A$ function that we have found
exists also away from criticality, which, in the language of this paper, would be away from the renormalization group fixed points of the $B$-functions, $B^I=0$.
The natural prospect of this idea would be to verify that it is possible to match $A$ and $G_{IJ}$ with $\tilde{F}$ and Zamolodchikov's metric on spheres. The diagrams required for the metric on spheres involve the same contractions of the coupling tensor that in this paper appear in the expansion of $G_{IJ}$, e.g., Eq.~\eqref{eq:metric1}
for the matching at leading order. The same computation has been done for the $\varphi^4$ model in Ref.~\cite{Pannell:2025ixz}, but in the case of the $\varphi^3$ there is the additional complication that to each order of the expansion the powers of the tensor coupling increase twice as fast.

The integrability of the renormalization group flow of $B^I$ can be conveniently expressed as a set of linear constraints that the coefficients of the $B$-functions themselves should satisfy.
Specifically, the $\varphi^3$ model is very constrained by integrability, especially if compared to $\varphi^4$. One explanation for this fact is, once more, that the polynomial degree of renormalization group functions grows twice as fast for $\varphi^3$, so the complexity of the three loop $\varphi^3$ is comparable to five/six loops $\varphi^4$, and we noticed this empirically when reproducing the results of Pannell and Stergiou of Ref.~\cite{Pannell:2024sia}. This statement is qualitative and must be taken with a grain of salt, in that higher loop diagrams still grow in complexity at the same rate in both models -- e.g., when looking at the higher order of transcendentality of the coefficients of the counterterms, which involve a larger field of numbers as coefficients. In fact, it would be interesting to explore in general if there is a relation between the number of constraints and the increase of complexity in the coefficients' field, given that the latter are linear relations.

In summary, there are in total $1+10$ constraints that the coefficients up to three loops must satisfy for the existence of $A$ in $d=6$. Our analysis reveals that there is one additional constraint among the three loop coefficients of $B^I$ for the gradient structure to hold in $d=6-\varepsilon$.

\smallskip

\paragraph*{Acknowledgments.} We are grateful to Andy Stergiou for initially suggesting this project and to Ian Jack and Hugh Osborn for patiently answering several questions on Ref.~\cite{Gracey:2015fia} and sharing the unpublished draft of Ref.~\cite{jack-osborn-unpublished-draft}. We are in debt with Hugh Osborn for pointing out an important mistake in a previous version of Eq.~\eqref{eq:curvature} which affected our original conclusions.

\appendix

\section{$\overline{\rm MS}$ coefficients and constraints}\label{sect:constraints}

In this appendix we include and occasionally repeat, order by order, the beta function coefficients in the $\overline{\rm MS}$ scheme. Plugging them in Eq.~\eqref{eq:gradient} we have solved the system for the coefficients of the $A$ function and the metric. They are explicitly included for the one and two loop cases, while in the three loop case we only include the $10$ constraint equations. All of the results of the loop expansion are the same as Refs.~\cite{Gracey:2015fia, jack-osborn-unpublished-draft}, but expressed in our notation.

At one loop the $\overline{\rm MS}$ coefficients of $B^{(1)}$ in \eqref{eq:beta1} are
\begin{align}\label{eq:MSbar-1loop-parameters}
 & c_{1,a}=\frac{1}{12}\,, && b_{1,a}=-1\,.
\end{align}
The values of the coefficients of the $A$-function are uniquely determined as
\begin{equation}\label{eq:Afun-values1}
 \begin{split}
  a_{1,a}=-\frac{1}{4}\,,
  \qquad a_{1,b} = \frac{1}{16}  \,.
 \end{split}
\end{equation}

At two loops the $\overline{\rm MS}$ values of $B^{(2)}$ in \eqref{eq:beta2} are 
\begin{align}\label{eq:MSbar-2loop-parameters}
 & c_{2,a}=\frac{1}{18}\,, && c_{2,b}=-\frac{11}{432}\,,
 \nonumber\\
 & b_{2,a}=-\frac{1}{4}\,, && b_{2,b}=\frac{7}{72}\,, && b_{2,c}=-\frac{1}{2}\,.
\end{align}
The coefficients of the $A$-function are solved as
\begin{align}\label{eq:Afun-values2}
  &
  a_{2,a}= -\frac{1}{12}\,,
  &&
  a_{2,b}= \frac{1}{6}-2\alpha_1 \,,
  &&
  a_{2,c}= \alpha_1 \,, \nonumber \\
  &
  a_{2,d}= -\frac{7}{576}-\frac{\alpha_1}{12}\,,
  &&
  a_{2,e}= \frac{1}{3456}- \frac{\alpha_1}{24} \,,
\end{align}
Similarly for the coefficients of $G_{IJ}$ are solved as
\begin{align}\label{eq:G-values2}
  &
  m_{1,a}= -\frac{7}{4}+12 \alpha_1 \,, 
  &&
  m_{1,b}= -\frac{7}{48} - \alpha_1 \,, \nonumber\\
  &
  m_{1,c}= \tilde{\alpha}_1-\alpha_1 \,, 
  &&
  m_{1,d}= \frac{1}{6}-\alpha_1 - \tilde{\alpha}_1 \,. 
\end{align}
The free parameters $\tilde{\alpha}_1$ and $\alpha_1$, discussed in Sect.~\ref{sect:further}, are unconstrained by integrability.
Notice that $m_{1,c}+m_{1,d}$ is independent on $\tilde{\alpha}_1$ as promised in the main text.

The two loop constraint from integrability given in the main text in Eq.~\eqref{eq:gradient-constraint-two-loop} is such that the values of the one loop coefficients of Eq.~\eqref{eq:MSbar-1loop-parameters} have been inserted.
In general we have
\begin{equation}\label{eq:gradient-constraint-two-loop}
 \begin{split}
  b_{1,a}c_{2,b}-2 c_{1,a}(b_{2,b}+c_{2,a})=0\,,
 \end{split}
\end{equation}
under the assumption that $b_{1,a}$ and $c_{1,a}$ are nonzero. Using \eqref{eq:MSbar-1loop-parameters} in \eqref{eq:gradient-constraint-two-loop} we find \eqref{eq:gradient-constraint-two-loop-previous-inserted} of the main text.
In the $\varepsilon$-expansion and at the same order, we can provide the $\varepsilon$ corrections to the coefficients of $A^{(1)}$ given in Eq.~\eqref{eq:Afun-values1}
\begin{equation}\label{eq:Afun-values1-epsilon}
 \begin{split}
  \tilde{a}'_{1,a}(0)=-\frac{7}{32} + \frac{3}{2} \alpha_1 \,,
  \qquad \tilde{a}'_{1,b}(0) = \frac{1}{384} -\frac{3}{8} \alpha_1 \,.
 \end{split}
\end{equation}
As expected, the corrections to $A$ at this order depend only on $\alpha_1$.

\begin{widetext}

In the $\overline{\rm MS}$ scheme, the three loop coefficients are
\begin{align}\label{eq:MSbar-3loop-parameters}
    & c_{3,a}= \frac{7}{864}\,, && c_{3,b}=\frac{71}{1728}\,, && c_{3,c}=-\frac{103}{10368}\,, && c_{3,d}= -\frac{1}{108}\,, && c_{3,e}= -\frac{121}{5184}\,,
    \nonumber\\
    & c_{3,f}= \frac{7}{96} - \frac{\zeta_3}{24}\,, && c_{3,g}= \frac{23}{62208} \,, && c_{3,h}= \frac{103}{7776}\,,  && c_{3,i}= -\frac{13}{31104}\,, && s_{3,a}= -\frac{137}{10368}\,,
    \nonumber\\
    & b_{3,a}= -\frac{3}{8}\,, && b_{3,b}= \frac{1}{8} \,, && b_{3,c}= \frac{5}{16} \,, && b_{3,d}=-\frac{47}{864} \,, && b_{3,e}=-\frac{47}{864} \,,
    \\
    & b_{3,f}= \frac{23}{288} \,, && b_{3,g}= \frac{5}{54}\,,  && b_{3,h}= \frac{11}{216}\,,  && b_{3,i}= -\frac{19}{324} \,, && b_{3,j}=\frac{11}{1728}\,,
    \nonumber\\
    & b_{3,k}= \frac{11}{1728} \,, && b_{3,l}= \frac{11}{144} \,, && b_{3,m}=-\frac{1}{16}\,,  && b_{3,n}= \frac{1}{3}-\zeta_3 \,, && b_{3,o}=-\frac{29}{48}+\frac{\zeta_3 }{2} \,,
    \nonumber\\
    & b_{3,p}=-1 \,, && b_{3,q}= -\frac{23}{24}+\zeta_3 \,,&&  &&  && 
    \nonumber
\end{align}
where $\zeta_3 = \zeta(3)$ and $\zeta$ is the Riemann $\zeta$-function.
We have included the $\overline{\rm MS}$ value of $s_{3,a}$ that has been determined indirectly in Ref.~\cite{Gracey:2015fia} (using the last equation of the system \eqref{eq:gradient-constraint-three-loop-previous-inserted} given below) and directly in Ref.~\cite{jack-osborn-unpublished-draft}.
The system of $10$ constraints becomes, inserting the $\overline{\rm MS}$ coefficients at one and two loops,
\begin{equation}\label{eq:gradient-constraint-three-loop-previous-inserted}
    \begin{split}
        &b_{3,b} + 2\, b_{3,c} + 12\, b_{3,d} + 24\, b_{3,e} + 12\, b_{3,f} + 144\, b_{3,l} + 72\, b_{3,m} + 72\, c_{3,e} + 864\, c_{3,i} + \frac{11}{9} - 3 \left(b_{3,a} + 4 \left(b_{3,g} + 72\, c_{3,g}\right)\right) = 0 \\
        &b_{3,b} + 12\, b_{3,h} - b_{3,a} - 12\, b_{3,g} = 0 \\
        &12\, b_{3,n} + b_{3,p} - 2\, b_{3,q} - \frac{7}{6} = 0 \\
        &b_{3,c} + 12\, b_{3,d} + 6\, b_{3,e} + 72\, b_{3,m} + \frac{7}{18} - b_{3,a} - 6\, b_{3,g} = 0 \\
        &b_{3,b} + 12\, b_{3,d} + 12\, b_{3,e} + 144\, b_{3,m} + 12\, c_{3,a} + 144\, c_{3,c} + \frac{11}{9} - b_{3,a} = 0 \\
        &b_{3,o} + \frac{1}{3} - b_{3,q} - 12\, c_{3,f} = 0 \\
        &b_{3,b} + 12\, b_{3,e} + 144\, b_{3,l} + 6\, c_{3,a} + 72\, c_{3,d} + 72\, c_{3,e} + 864\, c_{3,i} + 72\, s_{3,a} - b_{3,a} - 72\, b_{3,i} - \frac{11}{8} = 0 \\
        &b_{3,g} + 2\, c_{3,b} + 12\, c_{3,d} + 12\, s_{3,a} - b_{3,e} - 12\, b_{3,m} - c_{3,a} - 24\, c_{3,c} - \frac{49}{432} = 0 \\
        &b_{3,b} + 6\, b_{3,e} + 6\, b_{3,g} + 12\, c_{3,b} + 144\, s_{3,a} + 1 - b_{3,a} - 864\, c_{3,g} = 0 \\
        &c_{3,e} + 6\, c_{3,h} + 2\, s_{3,a} - \frac{77}{2592} = 0
        \,,
    \end{split}
\end{equation}
which is equivalent to that of Ref.~\cite{Gracey:2015fia} and satisfied by the $\overline{\rm MS}$ values given in Eq.~\eqref{eq:MSbar-3loop-parameters}.
The general form of the above constraints in which the coefficients of the previous loops are not inserted is
\begin{equation}\label{eq:gradient-constraint-three-loop}
    \begin{split}
        &
        c_{1,a} b_{3,q}-b_{1,a} b_{3,l}+2 b_{2,b} b_{2,c}-2 c_{1,a} b_{3,o}=0
        \\
        &
        b_{1,a} b_{3,i}-2 b_{1,a} s_{3,a}-2 b_{2,b}^2-2 c_{1,a} b_{3,f}=0
        \\
        &
        b_{3,a} c_{1,a}-b_{3,b} c_{1,a}-b_{1,a} b_{3,g}+b_{1,a} b_{3,h}=0
        \\
        &
        2 c_{1,a} c_{3,e}+4 c_{1,a} s_{3,a}+2 b_{2,b} c_{2,b} -b_{1,a} c_{3,h}=0
        \\
        &
        2 b_{1,a} c_{1,a} c_{3,f}+2 c_{1,a}^2 b_{3,m}-2 c_{1,a}^2 b_{3,o}+2 b_{2,b} c_{1,a} b_{2,c}-b_{1,a} b_{2,c} c_{2,b}=0
        \\
        &
        b_{1,a}^2 c_{3,g}-c_{1,a}^2 b_{3,e}-3 c_{1,a}^2 b_{3,g}+2 c_{1,a}^2 b_{3,h}+2b_{1,a}c_{1,a} s_{3,a}-b_{2,b} b_{1,a} c_{2,b}+b_{2,b}^2 c_{1,a}+b_{2,a}c_{1,a}c_{2,b}-2 c_{1,a}^2 c_{3,b}=0
        \\
        &
        2 c_{1,a}^2 b_{3,d}-2 b_{1,a} c_{1,a} c_{3,d}+4 c_{1,a}^2 b_{3,g}-2 c_{1,a}^2b_{3,h}-2 b_{1,a} c_{1,a} s_{3,a}-2 b_{2,b}^2 c_{1,a}+b_{2,b} b_{1,a} c_{2,b}-2b_{2,a} c_{1,a} c_{2,b}
        \\
        & \qquad +4 c_{1,a}^2 c_{3,b}+2 c_{3,c} b_{1,a} c_{1,a}=0
        \\
        &
        2 b_{1,a} c_{1,a}^2 b_{3,g}-b_{1,a}^2 c_{1,a} c_{3,d}-b_{1,a}^2 c_{1,a} s_{3,a}-2 b_{3,b} c_{1,a}^3+2 c_{1,a}^3 b_{3,c}+2 b_{2,b} b_{2,a} c_{1,a}^2+b_{1,a} c_{3,a} c_{1,a}^2+2 b_{1,a} c_{1,a}^2 c_{3,b}
        \\
        &\qquad -2 b_{2,b}^2 b_{1,a} c_{1,a}-2 b_{1,a} b_{2,a} c_{1,a} c_{2,b}+b_{2,b} b_{1,a}^2 c_{2,b}=0
        \\
        &
        2 b_{1,a} c_{1,a}^2 c_{3,d}+2 b_{1,a} c_{1,a}^2 c_{3,e}-4 c_{1,a}^3 b_{3,f}+8c_{1,a}^3 b_{3,g}-4 b_{1,a}^2 c_{1,a} c_{3,g}-4 c_{1,a}^3 b_{3,h}-2 b_{1,a}^2c_{1,a} c_{3,i}+4 b_{1,a} c_{1,a}^2 b_{3,j}
        \\
        &\qquad -10 b_{1,a} c_{1,a}^2 s_{3,a}+8c_{1,a}^3 c_{3,b}-8 b_{2,b}^2 c_{1,a}^2-2 b_{2,a} c_{1,a}^2 c_{2,b}+8 b_{2,b}b_{1,a} c_{1,a} c_{2,b}-b_{1,a}^2 c_{2,b}^2-2 c_{3,a} c_{1,a}^3=0
        \\
        &
        4 c_{1,a}^2 b_{3,g}-b_{1,a} c_{1,a} c_{3,d}-b_{1,a}^2 c_{3,g}-2c_{1,a}^2 b_{3,h}+b_{1,a} c_{1,a} b_{3,k}-3 b_{1,a} c_{1,a} s_{3,a}+b_{2,b}b_{1,a} c_{2,b}+2 c_{3,c} b_{1,a} c_{1,a}-2 b_{2,b}^2 c_{1,a}
        \\
        &\qquad -b_{2,a} c_{1,a}c_{2,b}+4 c_{1,a}^2 c_{3,b}-c_{1,a}^2 c_{3,a}=0 \,.
    \end{split}
\end{equation}
In the $\varepsilon$-expansion, eliminating $\tilde{a}'_{2,x}$ and $\tilde{m}'_{1,x}$ coefficients, we have reported the constraint Eq.~\eqref{eq:gradient-constraint-NLO-new-previous-inserted} in the main text inserting the coefficients of the previous orders. It can be simplified further using Eqs.\ \eqref{eq:gradient-constraint-three-loop-previous-inserted}
to the form
\begin{equation}\label{eq:gradient-constraint-NLO-new-previous-inserted-simpler}
\begin{split}
 &
 216 (b_{3,d}-b_{3,e}+ b_{2,f}-c_{3,a}) +18 (b_{3,a}- b_{3,b})
 -2592 \, c_{3,e}-7776 \, c_{3,f} +47=0
 \,.
\end{split}
\end{equation}
The general form of Eq.~\eqref{eq:gradient-constraint-NLO-new-previous-inserted} in which the previous loops have not been inserted is instead
\begin{equation}\label{eq:gradient-constraint-NLO-new}
\begin{split}
 &
 4 b_{1,a} c_{1,a}^2 c_{3,d}-2 b_{1,a} c_{1,a}^2 c_{3,e}+4 c_{1,a}^3 b_{3,f}-16 c_{1,a}^3 b_{3,g}+4
   b_{1,a}^2 c_{1,a} c_{3,g}+8 c_{1,a}^3 b_{3,h}+16 b_{1,a} c_{1,a}^2 s_{3,a}-16 c_{1,a}^3
   c_{3,b}
   \\&
   +12 b_{2,b}^2 c_{1,a}^2+4 b_{2,a} c_{1,a}^2 c_{2,b}-4 c_{3,c} b_{1,a} c_{1,a}^2-10
   b_{2,b} b_{1,a} c_{1,a} c_{2,b}+4 c_{3,a} c_{1,a}^3+b_{1,a}^2 c_{2,b}^2=0
 \,,
\end{split}
\end{equation}
and reduces to Eq.~\eqref{eq:gradient-constraint-NLO-new-previous-inserted}
when the values \eqref{eq:MSbar-1loop-parameters} and \eqref{eq:MSbar-2loop-parameters}
are used. Notice that the general constraint is now quartic in the coefficients, however, using \eqref{eq:gradient-constraint-two-loop} on the last term we could simplify it to a cubic equation.


%
\end{widetext}

%


\end{document}